\documentclass[11pt]{article}

\usepackage{graphicx}%% LaTeX Preamble - Common packages

\usepackage[utf8]{inputenc} % Any characters can be typed directly from the keyboard, eg éçñ
\usepackage{textcomp} % provide lots of new symbols
\usepackage{epstopdf} % to include .eps graphics files with pdfLaTeX
\usepackage{flafter}  % Don't place floats before their definition
\usepackage[utf8]{inputenc} % set input encoding (not needed with XeLaTeX)
\usepackage{amsthm,amssymb} 
\usepackage{amsfonts} 
\usepackage{amsmath}

\usepackage{bm}
\usepackage{color}
\usepackage{titlesec}
\usepackage[T1]{fontenc}
\usepackage{wrapfig}
\DeclareGraphicsExtensions{.pdf,.png,.jpg,.eps}
\usepackage{amsmath,amssymb}  % Better maths support & more symbols
\usepackage{bm}  % Define \bm{} to use bold math fonts

\usepackage{amscd}
\usepackage{eucal}
\usepackage{amssymb}

\usepackage{epic}
\usepackage{color}

\usepackage[pdftex,bookmarks,colorlinks,breaklinks]{hyperref}  % PDF hyperlinks, with coloured links
\usepackage{memhfixc}  % remove conflict between the memoir class & hyperref
\usepackage{pdfsync}  % enable tex source and pdf output syncronicity

\titleformat{\subsection}[runin]
{\normalfont\large\bfseries}{\thesubsection}{1em}{}
\newcommand{\p}{\partial}

\newcommand{\C}{{\mathbb C}}
\newcommand{\B}{{\mathbb B}}
\newcommand{\A}{{\mathbb A}}

\newtheorem{theorem}{Theorem}[section]

\newtheorem{cor}{Corollary}[section]

\begin{document}
\title{Nonlinear eigenvalue approximation for compact operators.}
\author{Shari Moskow
\thanks{Department of Mathematics, Drexel University, Philadelphia, PA 19104, USA, 
(moskow@math.drexel.edu).}}

\date{\today}

\maketitle

\abstract{In \cite{Os} a general spectral approximation theory was developed for compact operators on a Banach space which does not require that the operators be self-adjoint and also provides a first order correction term. Here we extend some of the results of that paper to nonlinear eigenvalue problems. We present examples of its application that arise in electromagnetics. }

\section{Introduction}

Frequently spectral perturbation problems arise in the study of electromagnetics and composite materials where the dependence on the spectral parameter is nonlinear. Here we state and prove some results which can be applied to both asymptotic and numerical approximations in those contexts. The general problem is as follows. Assume we have a set of compact linear operators $T_h(\lambda)$ where $\lambda$ is a spectral parameter and $h$ is some (w.l.o.g. small) perturbative parameter, and that we want to find a nontrivial pair $( u_h,\lambda_h)$ such that 
\begin{equation} \label{nolineig} \lambda_h T_h(\lambda_h) u_h=u_h. \end{equation}
Such a pair is what we refer to as a nonlinear eigenpair.
  We assume we have also some limiting problem
\begin{equation} \label{nolineiglim} \lambda_0 T_0(\lambda_0) u_0= u_0 \end{equation}
whose solutions are much simpler. The limiting problem could, for example,  correspond to a background problem where solutions are known, or a lower dimensional or homogeneous problem whose solutions are far easier to compute. To understand the behavior of $\lambda_h$, it is useful to have an expansion
\begin{equation}\label{eigasymp}\lambda_h = \lambda_0 + h\lambda^{(1)} +o(h) , \end{equation}
where the expression for $\lambda^{(1)}$ is as explicit as possible, and depends only on solutions to (\ref{nolineiglim}). We refer to $\lambda^{(1)}$ as the eigenvalue correction.

Note that in this class of nonlinear spectral problem, we are looking for $\lambda$ such that
 $ S(\lambda ) =\lambda T(\lambda)- I$ has a nontrivial null space, where unlike standard eigenvalue problems, $S$ depends nonlinearly on $\lambda$. The nonlinearity is only in $\lambda$, that is, we assume here that $T(\lambda)$ is itself a linear operator.  
The study of nonlinear eigenvalue problems, quadratic eigenvalue problems and operator pencils is not new, and we refer to \cite{Markus},\cite{chanillo},\cite{Ne}, as examples of a vast literature on the subject. However, this author is unaware of results on formulas for first order corrections to nonlinear eigenvalues for general Banach space operator perturbations.

\section{Background: Linear theory}
There is an established and expansive theory for linear eigenvalue perturbations of compact operators, in particular if they are self adjoint \cite{Ka}. 
%Of course it is well known that under certain conditions of convergence of the operators one has that the eigenvalues also converge, see for example the book \cite{Ka}, and that the error will be at least of the same order as that of the operator convergence. However, to obtain higher order approximations to be used for inversion, one would like to have the next order term, or correction.
To derive the correction for nonlinear eigenvalues, we begin with a {\it linear} eigenvalue correction theorem which is a restatement  of Theorem 3 of \cite{Os}.  This theorem does not require that the operators be self-adjoint, and within the proof provides an explicit formula for the correction (noticed in \cite{Vo}), which we state here. Suppose $X$ is a Banach space and $K_n: X\rightarrow X$ is a sequence
of compact linear operators such that $K_n\rightarrow K$ pointwise (i.e. $\forall f\in X$, $K_n f\rightarrow Kf$ in norm). Assume also that the sequence $\{ K_n \}$ is collectively compact, meaning that the set $\{ K_n f\ | \  \| f\|\leq 1, n=1,2,\ldots \}$ has compact closure. We also suppose that $K_n^*\rightarrow K^*$ pointwise and $\{ K_n^*\} $ is also collectively compact. (Note that all of the above conditions are met when $K_n\rightarrow K$ in the operator norm.)  Let
$\mu$ be a nonzero eigenvalue of $K$ of algebraic multiplicity $m$. It
is well known that for $n$ large enough, there exist $m$ eigenvalues
of $K_n$, $\mu_1^n,\ldots \mu_m^n$ (counted according to algebraic
multiplicity) such that $\mu_j^n\rightarrow\mu$ as
$n\rightarrow\infty$, for each $1\leq j\leq m$. Let $E$ be the spectral projection onto the generalized eigenspace of $T$ corresponding to eigenvalue $\mu$. The space $X$ can be decomposed in terms of the range and null space of $E$: $X=R(E)\oplus N(E)$. Elements in $\phi^*\in R(E)^*$ therefore can be extended to act on all of $X$ by initial projection onto $R(E)$, that is, $\phi^* f=\phi^* Ef $. 

\begin{theorem}[Osborn] [Linear Eigenvalue Corrections ] \label{osthm} 
Let $\phi_1,\phi_2,\ldots \phi_m$ be a normalized basis for $R(E)$, and let $\phi_1^*,\ldots \phi_m^* $ denote the corresponding dual basis for $R(E^*)$.
Then there exists a constant $C$ such that
 \begin{displaymath}
\left| \mu-{1\over{m}}\sum_{j=1}^m\mu_j^n -{1\over{m}}\sum_{j=1}^m \langle
(K-K_n)\phi_j,\phi^*_j\rangle \right| \leq C\|(K-K_n)|_{R(E)} \|\cdot \|
(K^*-K_n^*)|_{R(E^*)}\| .
\end{displaymath} 
\end{theorem}
Now suppose the operators depend on a continuous parameter $h$, and $K_h\rightarrow K$ in norm for example, and they differ by order $h$. 
Note that the right hand side is of higher order and the correction term for the average of the perturbed eigenvalues is merely 
$$ {1\over{m}}\sum_{j=1}^m \langle (K_h-K)\phi_j,\phi^*_j\rangle$$
so that if one has $K_h\approx K + h K^{(1)} $ for $K^{(1)}$ the operator correction, this yields the formula ${1\over{m}}\sum_{j=1}^m\mu_j^h \approx \mu +h\mu^{(1)}$
where
$\mu^{(1)}= {1\over{m}}\sum_{j=1}^m \langle K^{(1)} \phi_j,\phi^*_j\rangle.$ Of course for a simple eigenpair $(\mu,\phi)$ this means that
$$\mu_h = \mu + h \langle K^{(1)} \phi,\phi^*\rangle +O(h^2).$$

\section{Convergence of nonlinear eigenvalues}

One expects convergence of the nonlinear eigenvalues due to the analytic Fredholm theory. Here we include a statement of convergence and its proof for completeness.
Define the modified resolvent type operator valued functions on $\mathbb{C}$
$$R_h(\lambda)=(I-\lambda T_h(\lambda))^{-1}$$ and
$$R_0(\lambda)=(I-\lambda T_0(\lambda))^{-1}.$$ An important note is that if
$R_h(\lambda)$ does not exist as a bounded linear operator from $X$ to
itself, then $\lambda$ is a nonlinear eigenvalue of $T_h$.  This is because if
$R_h(\lambda)$ does not exist, then ${1/\lambda }$ is in the spectrum of the
compact operator $T_h(\lambda)$. Hence since $1/\lambda$ is nonzero, it must be an eigenvalue and
$(I-\lambda T_h(\lambda))$ must have nontrivial and finite dimensional null
space. The same argument holds for the limiting operator $R_0(\lambda)$.

In the following proposition, we show that the nonlinear eigenvalues converge to those of the unperturbed operator.

\begin{theorem} \label{convergeprop} Assume that $\lambda_0$ is a nonlinear  eigenvalue of $T_0$, and that $R_0$ and $R_h$ are meromorphic in some region $U$ of $\mathbb{C}$ containing $\lambda_0$. Assume also that for any $\lambda\in U$, $T_h(\lambda)\rightarrow T_0(\lambda)$ in norm. Then for any ball $B$ around $\lambda_0$, there exists $h_0 >0$  such that $T_h$ has a nonlinear eigenvalue in $B$ for all $h<h_0$. Conversely, if $\{ \lambda_h\}$ is a sequence of nonlinear eigenvalues of $T_h$ that converges as $h\rightarrow 0$, the limit is a nonlinear eigenvalue of $T_0$. \end{theorem}
{\it Proof} \ \ \ \ \  Since $U$ is open and $R_0$ meromorphic, we can choose $B$, a ball around $\lambda_0$ such that $T_0$ has no other nonlinear  eigenvalues in $\overline{B}$.  
We will use a well known result about the inverses of
perturbed operators, see for example \cite{Ka} p. 31: If $S-T=A$ and $T^{-1}$ exists, then for $\| A\| < {1\over{\| T^{-1}\| }}$, $S^{-1}$ exists and 
\begin{equation} \label{inversebound} \| S^{-1}-T^{-1}\| \leq {\| A\| \| T^{-1}\|^2 \over{1-\| A\| \| T^{-1}\| }}.\end{equation}
Apply this, with 
$$S=I-\lambda T_h(\lambda)$$
$$T=I-\lambda T_0(\lambda)$$ to get 
\begin{equation}\label{resolventerror}\| R_h(\lambda)-R_0(\lambda)\| \leq {\lambda \| T_0(\lambda)-T_h(\lambda)\| \| R_0(\lambda)\|^2\over{1-\lambda\|  T_0(\lambda)-T_h(\lambda)\| \| R_0(\lambda)\|}}.\end{equation}
Let $\Gamma=\p B$, positively oriented. By the choice of $B$, $\Gamma$ does not intersect with any poles of $R_0$, and $\lambda_0$ is the only pole of $R_0$ in the closed disk. Then $R_0(\lambda)$ is continuous with respect to $\lambda$ on $\Gamma$, hence 
$\| R_0(\lambda)\|$ is uniformly bounded for $\lambda$ on $\Gamma$. Using (\ref{resolventerror}),  we have that
$$R_h(\lambda)\rightarrow R_0(\lambda)$$
in norm as $h\rightarrow 0$, uniformly for $\lambda\in \Gamma$. This implies that the operator valued integral
$${1\over{2\pi i}}\int_\Gamma R_h(\lambda)d\lambda\rightarrow {1\over{2\pi i}}\int_\Gamma R_0(\lambda)d\lambda $$  in norm as $h\rightarrow 0$. 
More generally, the integrals
$${1\over{2\pi i}}\int_\Gamma (\lambda-\lambda_0)^\alpha R_h(\lambda)d\lambda\rightarrow {1\over{2\pi i}}\int_\Gamma (\lambda-\lambda_0)^\alpha R_0(\lambda)d\lambda $$ 
also converge for $\alpha$ a positive integer.
From the residue theorem, the integral 
$${1\over{2\pi i}}\int_\Gamma (\lambda-\lambda_0)^\alpha R_0(\lambda)d\lambda $$
gives us the coefficient of the $(\lambda-\lambda_0)^{-(\alpha+1)}$ term in the Laurent series expansion for $R_0(\lambda)$. Since $R_0$ has a pole at $\lambda_0$ and is meromorphic, this must be nonzero for some finite integer $\alpha\geq 0$.  Hence for that $\alpha$, the integrals 
$${1\over{2\pi i}}\int_\Gamma (\lambda-\lambda_0)^\alpha R_h(\lambda)d\lambda$$ must all be nonzero for $h$ small enough. This means that all $R_h$ must have at least one pole in $B$ for $h$ small enough. That is, for $h$ small enough, all $T_h$ have a nonlinear eigenvalue in $B$. This proves the first part of the statement of the proposition. For the converse, if $\lambda_0$ is not a nonlinear eigenvalue of $T_0$, then $R_0(\lambda)$ exists in some neighborhood of $\lambda_0$. The formula (\ref{resolventerror}) implies that $R_h(\lambda)$ also exists in that neighborhood for $h$ small enough. Hence the nonlinear eigenvalues of $T_h$ are bounded away from $\lambda_0$ for $h$ small enough.  \hfill{$\Box$}

Some remarks about the assumptions in this theorem:
\begin{itemize}
\item If the operator functions $I-\lambda T_h(\lambda)$ and 
$I-\lambda T_0(\lambda)$ are analytic in some region $U$, then this combined with the fact that the $T$'s are compact, means that the inverses are meromorphic. % reference? What we need is that the poles are isolated. 

\item If $\lambda_0$ is a nonlinear eigenvalue of $T_0$, then the classical resolvent of $T_0(\lambda_0)$, given by $ (zI- T_0(\lambda_0))^{-1}$, automatically has nonzero residue at $z={1\over{\lambda_0}}$; its residue is the projection onto the generalized eigenspace \cite{Ka}. Using arguments as above, one can show that the coefficient operators in the Laurent series expansions for $R_h$ must converge to those of $R_0$.  However, it is not clear how these coefficients relate to the nonlinear eigenspaces.

\end{itemize}

\section{Nonlinear eigenvalue corrections}
Now assume we have a series of problems of the form (\ref{nolineig}),(\ref{nolineiglim}). For the case of resonances for the Helmholtz equation we were able to extend and modify the linear eigenvalue correction theorem of Osborn to apply such a situation \cite{GoMoSa}. Using the ideas there, we will state and prove a general nonlinear eigenvalue correction theorem that works for simple eigenvalues, or for higher multiplicity if the perturbed nonlinear eigenvalue has multiplicity just as high. We remark that the following theorem does not apply at all for multiple eigenvalues in the general case.

If $\lambda_0$ is a nonlinear eigenvalue of $T_0$, then ${1\over{\lambda_0}}$ is a standard eigenvalue of $T_0(\lambda_0)$, with algebraic multiplicity $m$ and $E$ the projection onto the corresponding generalized eigenspace. We will say in this case that $\lambda_0$ has multiplicity $m$. As in the linear case, let $\{ \phi_j \}_{j=1,\ldots m},\{ \phi^*_j \}_{j=1,\ldots m} $ be normalized bases of the generalized eigenspace $R(E)$ and its dual space $R(E)^*$ respectively. Again $X=R(E)\oplus N(E)$, and elements in $\phi^*\in R(E)^*$  can be extended to act on all of $X$ by initial projection onto $R(E)$, that is, $\phi^* f=\phi^* Ef $. Here $X$ is again a Banach space, and $\langle f,g^* \rangle=g^*(f)$ represents the duality pairing for $f\in X, g^*\in X^*$.  

\begin{theorem}[Nonlinear Eigenvalue Corrections] \label{nonlinearosthm}  Let $\{ T_h(\lambda):X\rightarrow X \} $ be a set of compact linear operator valued functions of $\lambda$ which are analytic in a region $U$ of the complex plane, collectively compact for any $\lambda\in U$. Assume that $T_h(\lambda)\rightarrow T_0(\lambda)$ pointwise as $h\rightarrow 0$, uniformly for $\lambda\in U$.  Assume also that  $T^*_h(\lambda)\rightarrow T^*_0(\lambda)$ pointwise as $h\rightarrow 0$, and that $\{ T^*_h(\lambda)\} $ are collectively compact, uniformly for $\lambda\in U$. Let $\lambda_0\neq 0, \lambda_0\in U$ be a nonlinear eigenvalue (\ref{nolineiglim}) of $T_0$ , of algebraic multiplicity $m$.   
Assume that there exists $h_0$ such that  for $h< h_0$ there exists  $\{ \lambda_h\} $  a nonlinear eigenvalue of $T_h$ of multiplicity m, such that  $\lambda_h\rightarrow \lambda_0$. Assume $B\subset U$ is a ball around $\lambda_0$ containing all $\lambda_h$ for $h<h_0$.
Let $DT_0(\lambda_0)$ be the derivative of $T_0$ with respect to $\lambda$ evaluated at $\lambda_0$.  Then  if 
\begin{equation}\label{DTcond}{ \lambda_0^2\over{m}} \sum_{j=1}^m \langle DT_0(\lambda_0) \phi_j,\phi_j^*\rangle\neq -1 ,\end{equation} we have the following formula
 \begin{multline} \label{nonlinearcor}
\lambda_h=\lambda_0 + {{\lambda_0^2\over{m}}\sum_{j=1}^m\langle (T_0(\lambda_0)-T_h(\lambda_0))\phi_j, \phi_j^*\rangle \over {1+{\lambda_0^2\over{m}}\sum_{j=1}^m\langle DT_0 (\lambda_0) \phi_j,\phi_j^* \rangle}}  \\ + O\left( \sup_{\lambda\in B}  \| (T_h(\lambda)- T_0(\lambda))|_{R(E)} \| \| (T^*_h(\lambda)- T^*_0(\lambda))|_{R(E)^*}\|  \right).
\end{multline}  
\end{theorem}

{\it Proof}  \ \ \ \ \  Note that
$$\lambda_h T_h(\lambda_h) u_h=u_h$$
and
$$\lambda_0 T_0(\lambda_0) u_0=u_0,$$
that is, ${1\over{\lambda_h}}$ is an eigenvalue of $T_h(\lambda_h )$ and
${1\over{\lambda_0}}$ is an eigenvalue of $T_0(\lambda_0)$. Also, by assumption we know that
$$T_h(\lambda_h)\rightarrow T_0(\lambda_0)$$ pointwise, and the sequence is collectively compact; likewise for the adjoints. So, what we have are the eigenvalues of a
convergent sequence of compact operators, $\{ {1\over{\lambda_h}}, T_h(\lambda_h)\} $ converging to $\{ {1\over{\lambda_0}}, T_0(\lambda_0) \} $, so we now apply Theorem \ref{osthm}. Since ${1\over{\lambda_h}}$ has multiplicity $m$, all of the eigenvalues of $T_h(\lambda_h)$ are equal to ${1\over{\lambda_h}}$ if $h$ is small enough. (We remark that if this is not the case, the desired nonlinear eigenvalue must get averaged with the other {\it linear} eigenvalues of $T_h(\lambda_h)$, which are not of interest to us here.)  Hence this theorem yields
\begin{multline} 
\left| {1\over{\lambda_0}}-{1\over{\lambda_h}}
- {1\over{m}}\sum_{j=1}^m \langle
(T_0(\lambda_0)-T_h(\lambda_h))\phi_j,\phi_j^*\rangle \right|
\\  \leq \| (T_h(\lambda_h)- T_0(\lambda_0))|_{R(E)} \| \| (T^*_h(\lambda_h)- T^*_0(\lambda_0))|_{R(E)^*} \| . \label{osborn2}
\end{multline} 
Since $R(E)$ is finite dimensional,  $$ \| (T_h(\lambda)- T_0(\lambda))|_{R(E)} \| \leq c(h) $$ where $c(h)\rightarrow 0$ as $h\rightarrow 0$ and is independent of $\lambda\in U$. Similarly, we have $$ \| (T^*_h(\lambda)- T^*_0(\lambda))|_{R(E)^*} \| \leq c^*(h) $$  where $c^*(h)\rightarrow 0$ as $h\rightarrow 0$.  (Note that these rates need not necessarily be the same.)
From the regularity of $T_0$ with respect to $\lambda$ and the assumptions of this theorem, there exists $C$ independent of $h$ and $\lambda\in U$ such that  
$$\| T_0(\lambda_0)-T_h(\lambda_h)|_{R(E)}\| \leq C( c(h) + |\lambda_h-\lambda_0|  ) $$
and likewise $$ \|
T^*_0(\lambda_0)-T_h^*(\lambda_h)|_{R(E)^*} \|\leq C( c^*(h) + |\lambda_h-\lambda_0|  ) .$$ 
Inserting this into (\ref{osborn2}) we have
\begin{multline} 
\left| {1\over{\lambda_0}}-{1\over{\lambda_h}}
- {1\over{m}}\sum_{j=1}^m \langle
(T_0(\lambda_0)-T_h(\lambda_h))\phi_j,\phi_j^*\rangle \right|
\\  \leq C( c(h) + |\lambda_h-\lambda_0|  )\cdot( c^*(h) + |\lambda_h-\lambda_0|  ) .   \label{osborn3}
\end{multline} 
 If we
multiply everything by $\lambda_0\lambda_h$,
\begin{equation} \left|\lambda_h-\lambda_0-{\lambda_0\lambda_h\over{m}} \sum_{j=1}^m\langle
(T_0(\lambda_0)-T_h(\lambda_h))\phi_j,\phi_j^*\rangle\right|\leq C( c(h) + |\lambda_h-\lambda_0|  )\cdot( c^*(h) + |\lambda_h-\lambda_0|  )\nonumber \end{equation} which we
manipulate to get
\begin{multline} \nonumber \lambda_h=\lambda_0 +{\lambda_0^2\over{m}}\sum_{j=1}^m\langle
(T_0(\lambda_0)-T_h(\lambda_h))\phi_j,\phi_j^*\rangle 
+{\lambda_0\over{m}}(\lambda_h-\lambda_0)\sum_{j=1}^m\langle
(T_0(\lambda_0)-T_h(\lambda_h))\phi_j,\phi_j^* \rangle \\ +O\left(
(c(h)+|\lambda_h-\lambda_0|)\cdot (c^*(h)+|\lambda_h-\lambda_0|)\right). \nonumber \end{multline} 
Now, we again see that the third term on the right hand side is bounded by the error term, and hence
 \begin{multline}  \lambda_h=\lambda_0 +{\lambda_0^2\over{m}}\sum_{j=1}^m\langle
(T_0(\lambda_0)-T_h(\lambda_h))\phi_j,\phi_j^*\rangle 
+O\left(
(c(h)+|\lambda_h-\lambda_0|)\cdot (c^*(h)+|\lambda_h-\lambda_0|)\right). \label{eexpand} \end{multline} 
Now, since the
correction term above depends on $\lambda_h$, we need to expand the
term further.  We can write
\begin{equation}
T_0(\lambda_0) -T_h(\lambda_h)= (T_0(\lambda_0)
-T_h(\lambda_0))+(T_h(\lambda_0) -T_h(\lambda_h)) \label{Tdivide2}
\end{equation}
and compute using the regularity with respect to $\lambda$,
\begin{equation}\langle (T_h(\lambda_0) -T_h(\lambda_h))\phi_j,\phi_j^*\rangle =(\lambda_0-\lambda_h) \langle DT_h (\lambda_0)\phi_j,\phi_j^*\rangle +O(|\lambda_0-\lambda_h|^2) \nonumber
\end{equation}
where $DT_h(\lambda_0)$ is the derivative with respect to $\lambda$ of $T_h$ evaluated at $\lambda_0$. 
Since the pointwise convergence of $T_h$ is uniform with respect to $\lambda$, we have that
$$\langle (DT_h (\lambda_0)-DT_0(\lambda_0) )\phi_j,\phi_j^*\rangle \leq C c(h) $$
which yields 
\begin{multline}\langle (T_h(\lambda_0) -T_h(\lambda_h))\phi_j,\phi_j^*\rangle =(\lambda_0-\lambda_h) \langle DT_0 (\lambda_0)\phi_j,\phi_j^*\rangle \\ +O\left(
(c(h)+|\lambda_h-\lambda_0|)\cdot (c^*(h)+|\lambda_h-\lambda_0|)\right) \label{dest}
\end{multline}

Combining
(\ref{eexpand}), (\ref{Tdivide2}) , and (\ref{dest}), we obtain
\begin{multline} \lambda_h=\lambda_0+{\lambda_0^2\over{m}}\sum_{j=1}^m \langle
(T_0(\lambda_0)-T_h(\lambda_0))\phi_j,\phi_j^*\rangle
-{\lambda_0^2\over{m}} (\lambda_h-\lambda_0)\sum_{j=1}^m\langle DT_0(\lambda_0)\phi_j,\phi_j^* \rangle \\ +
O\left((c(h)+|\lambda_h-\lambda_0|)\cdot (c^*(h)+|\lambda_h-\lambda_0|) \right) \nonumber .\end{multline}  
We now collect terms for $(\lambda_h-\lambda_0)$ so that 
\begin{multline} (\lambda_h-\lambda_0)\left(1+{\lambda_0^2\over{m}}\sum_{j=1}^m\langle DT_0(\lambda_0)\phi_j,\phi_j^* \rangle \right) ={\lambda_0^2\over{m}}\sum_{j=1}^m \langle
(T_0(\lambda_0)-T_h(\lambda_0))\phi_j,\phi_j^*\rangle \\ +
O\left((c(h)+|\lambda_h-\lambda_0|)\cdot (c^*(h)+|\lambda_h-\lambda_0|) \right). \end{multline} 
At this point we need to use the assumption (\ref{DTcond}) to obtain 
\begin{multline}\label{finalest}  \lambda_h=\lambda_0 + {{\lambda_0^2\over{m}}\sum_{j=1}^m\langle (T_0(\lambda_0)-T_h(\lambda_0))\phi_j, \phi_j^*\rangle \over {1+{\lambda_0^2\over{m}}\sum_{j=1}^m\langle DT_0 (\lambda_0) \phi_j,\phi_j^* \rangle}} \\ + O\left((c(h)+|\lambda_h-\lambda_0|)\cdot (c^*(h)+|\lambda_h-\lambda_0|) \right) \end{multline}
By the looking at either the operators or their adjoints, we must have that 
$$\langle (T_0(\lambda_0)-T_h(\lambda_0))\phi_j,\phi_j^*\rangle \leq \min{\{ c(h),c^*(h)\} } ,$$
which implies that  
 $$\lambda_h-\lambda_0=O(\min{\{ c(h),c^*(h)\} } ) +O\left((c(h)+|\lambda_h-\lambda_0|)\cdot (c^*(h)+|\lambda_h-\lambda_0|) \right) .$$
 Since we assume that $\lambda_h-\lambda_0
\rightarrow 0$, this can only hold if $$\lambda_h-\lambda_0=O(\min{\{ c(h),c^*(h)\} }).$$ Inserting this into (\ref{finalest}) 
completes the proof.\hfill{$\Box$}

The following simplified version of the above theorem is applicable in many situations. 
\begin{cor}\label{nonlinearcor}  Let $\{ T_h(\lambda):X\rightarrow X \} $ be a set of compact linear operator valued functions of $\lambda$ which are analytic in a region $U$ of the complex plane, such that $T_h(\lambda)\rightarrow T_0(\lambda)$ in norm as $h\rightarrow 0$ uniformly for $\lambda\in U$.   Let $\lambda_0\neq 0, \lambda_0\in U $ be a simple nonlinear eigenvalue (\ref{nolineiglim}) of $T_0$,
define $DT_0(\lambda_0)$ to be the derivative of $T_0$ with respect to $\lambda$ evaluated at $\lambda_0$, and let $\phi$ be the normalized eigenfunction and $\phi^*$ its dual.  Then for any $h$ small enough there exists $\lambda_h$ a simple nonlinear eigenvalue of $T_h$ ,  such that  if 
\begin{equation}\label{DTcond}{ \lambda_0^2}  \langle DT_0(\lambda_0) \phi,\phi^*\rangle\neq -1 ,\end{equation} we have the following formula
 \begin{multline} \label{nonlinearcorform}
\lambda_h=\lambda_0 + {{\lambda_0^2}\langle (T_0(\lambda_0)-T_h(\lambda_0))\phi, \phi^*\rangle \over {1+{\lambda_0^2}\langle DT_0 (\lambda_0) \phi,\phi^* \rangle}}  \\ + O\left(\sup_{\lambda\in U}  \| (T_h(\lambda)- T_0(\lambda))|_{R(E)} \| \| (T^*_h(\lambda)- T^*_0(\lambda))|_{R(E)^*}\|  \right).
\end{multline}   
\end{cor}
\section{Examples}
\subsection{Generalized eigenvalue problems}\label{gensub}
If one has a sequence of generalized eigenvalue problems of the form $$ Au=\lambda B u,$$ then if either $A$ or $B$ is invertible and the other compact, this case reduces to the linear case and one can use Theorem \ref{osthm}, assuming all conditions are met. However, if instead one has the form $$ (A +K)u=\lambda Bu$$ where $A$ is invertible and $K$,$B$ are compact, this does not reduce to a standard eigenvalue problem unless one knows that $(A+K)$ is also invertible. However, one can multiply by $A^{-1}$ to have
$$ u=( -A^{-1}K+\lambda A^{-1} B)u , $$ 
or
$$ u= \lambda ( -{1\over{\lambda}} A^{-1}K+ A^{-1} B) u.$$
Hence for perturbations of this problem, one can potentially apply Theorem \ref{nonlinearosthm} with $$T(\lambda) = -{1\over{\lambda}} A^{-1}K+ A^{-1} B . $$
\subsection{Transmission eigenvalues}
In classical scattering problems, the far field scattering operator is a measure of the difference between the free space solution and solution of the equation modeling the presence of a (bounded here) scatterer. 
Wave numbers for which the scattering operator has a nontrivial kernel yield  solutions to what is called the interior transmission eigenvalue problem for a given scatterer. Since these transmission eigenvalues are so closely related to non-scattering incident waves \cite{CakHad2}, \cite{sp}, it is clear they play an important role in inversion \cite{ccm07}, \cite{haddex}.  There has been quite a bit of progress made in these problems, including existence proofs \cite{cakginhad}  \cite{paisyl}  \cite{coltpaisyl}. See also the survey paper  \cite{CakHad2}.

 Let $D\subset {\mathbb R}^d$, $d\geq 2$ be a bounded connected region with smooth boundary $\partial D$ and  let $\nu$ denote the unit normal vector oriented outward to $D$. We consider a real valued function $n(x)$ defined in $D$, such that  and $n(x)\geq n_0>0$.  The transmission eigenvalue problem associated with $D$, $n$ are the values of $k$ for which the interior transmission problem
\begin{eqnarray}
&\Delta v+k^2v=0&\qquad \mbox{in}\; {D}\label{iit1}\\
&\Delta w+k^2n(x)w=0&\qquad \mbox{in}\; {D} \label{ids2}\\
&w=v \qquad
\displaystyle{\frac{\partial w}{\partial \nu}=\frac{\partial v}{\partial \nu}} &\qquad \mbox{on} \;\partial D\label{iit4}
\end{eqnarray}
has a nontrivial solution pair $(v, w)$. It was shown in \cite{cakginhad} under a fixed sign assumption on $n-1$ one can write the equivalent eigenvalue problem for $u=v-w\in H^2_0(D)$:
\begin{equation}\label{gr}
\left(\Delta+k^2n\right)\frac{1}{n-1}\left(\Delta +k^2\right)u=0.
\end{equation}
Note that this eigenvalue problem is not self adjoint and it is no longer linear, it is quadratic in $\lambda=k^2$. So its analysis is not covered by standard elliptic eigenvalue theory.  This can be multiplied out and written in the operator form \cite{cakginhad}
$$\A u+\, \lambda \B u+ \lambda^2 \C u=0$$
where $\A , \B , \C :\!H^2_0(D)\to H^2_0(D)$ are defined using Riesz representation
$$\left(\A u,v\right)_{H_0^2(D)}=\int_D\frac{1}{n -1}\Delta u \, \Delta\overline{v} \, \textrm{d}x,$$
$$\left(\B  u,v\right)_{H_0^2(D)}=\int_D\frac{1}{n -1}\left(\Delta u\,\overline v + n u\, \Delta\overline v \right)\,dx $$
$$ \left(\C  u, v\right)_{H_0^2(D)} = \int_D\frac{n}{n-1} u \, \overline{v} \, dx.$$
Here the $H^2_0$ inner product is the $L^2$ inner product of the Laplacians.  
Note that $\A $ is invertible and $\B, \C$ are compact. 
Imagine we are interested in how these transmission eigenvalues are perturbed by  material perturbations (i.e. perturbations in $n$ or $D$), or by numerical approximations (i.e. discretizations of $\A, \B, \C$). Since quadratic eigenvalue problems can be converted to 2$\times$2 linear problems,  Osborn's Theorem \ref{osthm} may be applied directly, and this approach was indeed used in \cite{CaMo}. However, depending on the type of perturbation, the use of a system may be quite inconvenient, or may not work at all (e.g. in \cite{CaMo} the formula does not work for complex eigenvalues).  The ability to use a nonlinear formulation leaves us a number of options. The simplest may be to invert $\A$, and use 
$$ T(\lambda)= -\A^{-1} \B -\lambda \A^{-1}\C .$$
If we have a simple nonlinear eigenvalue $\lambda_0$ and some perturbation indexed by $h$, then Corollary \ref{nonlinearcor} (assuming all hypotheses are met) yields 
\begin{multline} \label{nonlinearcorform}
\lambda_h=\lambda_0 + {{\lambda_0^2}\langle (\A_h^{-1} \B_h +\lambda_0 \A_h^{-1}\C_h - ( \A^{-1} \B +\lambda_0 \A^{-1}\C ))\phi, \phi^*\rangle \over {1+{\lambda_0^2}\langle \A^{-1}\C \phi,\phi^* \rangle}}  \\ + O(  \| (T_h(\lambda)- T_0(\lambda))|_{R(E)} \| \| (T^*_h(\lambda)- T^*_0(\lambda))|_{R(E)^*}\| .  )
\end{multline}   
where one expects the square of the norms to be asymptotically smaller than the correction term. Furthermore, when $n-1$ is sign changing, one will need to use the formulation in \cite{sy}, a generalized eigenvalue problem which for some situations can be converted to an compact nonlinear problem in the manner presented in subsection \ref{gensub}.

\subsection{Resonances}
 Assume we are interested in the propagation of scalar waves in free space, $x\in\mathbb{R}^d$, with the presence of  a scattering obstacle. The fields may obey the Helmholtz equation: 
$$\Delta u(x) + k^2 (1 +\eta(x) ) u=0 $$ 
where $u=u_{\bf i} + u_{\bf s}$ is divided into a given incident wave and scatterered field, and where the scattered field $u_{\bf s}$ satisfies the appropriate Sommerfeld radiation conditions at infinity. Here the scatterer is modeled by  $\eta(x)$, which has compact support contained in the compact set $D$.  Standard integration by parts yields the equivalent Lippmann-Schwinger formulation for the total field $u$,
$$u(x)=u_{\bf i}(x) + k^2 \int_{D} {\eta}(y)  G(x,y) u({y}) d{y} $$
where, in dimension three for example,  $$ G(x,y) = {1\over{4\pi}} {e^{ik|x-y|}\over{|x-y|}}$$ 
is the Helmholtz fundamental solution.
Unlike waves in a bounded domain, operators in free space with an obstacle scatterer do not have real eigenvalues and corresponding standing waves/modes. However, if one allows $\lambda= k^2$  to extend into the complex plane, you can have poles, or values of $\lambda$ where the above equation has nontrivial solutions for no incident wave. Although such solutions are nonphysical, if their imaginary part is small they are observed as resonances. In the time domain these solutions correspond to very slowly decaying modes. That is, the resonances are values of $\lambda$ for which
$$u(x)= \lambda \int_{D} {\eta}(y) {1\over{4\pi}} {e^{i\sqrt{\lambda}|x-y|}\over{|x-y|}} u({y}) d{y} $$
has nontrivial solutions $u$. This is a nonlinear eigenvalue problem where the operator
$$( T(\lambda)u)(x) =  \int_{D} {\eta}(y) {1\over{4\pi}} {e^{i\sqrt{\lambda}|x-y|}\over{|x-y|}} u({y}) d{y} $$
is analytic with respect to $\lambda$  away from the negative real axis. Perturbations in $\eta$ which correspond to material defects, or numerical discretizations of such an operator, can be handled by the theory presented here for the case of simple eigenvalues. Since the eigenvalue problem is not polynomial, converting to a linear system would require the use of an infinite system and would potentially be far more complicated. 

\section*{Acknowledgements}
The author was supported by the National Science Foundation under grant DMS-1108858.

%\bibliography{eigpaper} 
\bibliographystyle{plain} 

\end{document}